

\def\NI{\noindent}

\long\def\UN#1{$\underline{{\vphantom{\hbox{#1}}}\smash{\hbox{#1}}}$}
\magnification=\magstep 1
\overfullrule=0pt
\hfuzz=16pt
\voffset=0.0 true in
\vsize=8.8 true in
\def\NP{\vfil\eject}
\baselineskip 20pt
\parskip 6pt
\hoffset=0.1 true in
\hsize=6.3 true in
\nopagenumbers
\pageno=1
\footline={\hfil -- {\folio} -- \hfil}

\

\

\centerline{\UN{\bf Kinetic Roughening in Deposition with
Suppressed Screening}}

\vskip 0.4in

\centerline{\bf Peter Nielaba$^a$\
{\rm and}\ Vladimir Privman$^b$}

\vskip 0.2in

\NI\hang $^a${\sl Institut f\"ur Physik, Universit\"at Mainz,
Staudingerweg 7, D--55099 Mainz, Germany}

\NI\hang $^b${\sl Department of Physics, Clarkson University,
Potsdam, New York 13699--5820, USA}

\vskip 0.4in

\centerline{\bf ABSTRACT}

Models of irreversible surface deposition of $k$-mers on a linear
lattice, with screening suppressed by disallowing overhangs blocking
large gaps, are studied by extensive Monte Carlo simulations of the
temporal and size dependence of the growing interface width.
Despite earlier finding that for such models the deposit density
tends to increase away from the substrate, our numerical results
place them clearly within the standard KPZ universality class.

\vfill

\NI {\bf PACS numbers:}$\;$  68.10.Jy, 82.20.Wt

\

\NP

The standard, KPZ [1] model of kinetic roughening of growing
surfaces (reviewed, e.g., in [2-3]) yields the scaling prediction
for the interfacial width $W$ as a function of time, $T$, and
substrate size, $L$,

$$ W \simeq L^\zeta F \left( T L^{-z} \right) \;\; , \eqno(1) $$

\NI where for 1D substrates,

$$ \zeta_{\rm KPZ} = {1 \over 2} \;\; , \eqno(2) $$

$$ z_{\rm KPZ} = {3 \over 2} \;\; . \eqno(3) $$

\NI In fact, the value $\zeta =1/2$ is typical of 1D fluctuating
interfaces. However, (3) is characteristic of the KPZ
universality class. For instance, for stationary
fluctuating interfaces, (3) is replaced by $z=2$. These values
have been well established by numerical simulations and are
believed to be exact (in 1D).

In an interesting study [4], Krug and Meakin argued that to the
leading order, the KPZ fluctuations affect the growth rate by
introducing, in the average deposit density $\rho (h)$ at the
height $h$ away from the substrate, the term

$$ \Delta \rho_{\rm KPZ} \simeq \lambda h^{-2(1-\zeta )/z} \;\; .
\eqno(4)$$

\NI This expression applies for times $T$ large enough so that
the density has reached its limiting value at $h$, and assuming no
finite-$L$ effects, i.e., for infinite substrates. The
coefficient $\lambda >0$ is related to the nonlinear growth term is
the KPZ theory [1-3]. Specifically, in 1D, this contribution
suggests the coverage \UN{\it decreasing} \ to the limiting
large-$h$ value according to the power law $\Delta \rho \sim
h^{-2/3}$. The prediction (4) has been verified for several
ballistic deposition models in 1D and 2D [4-5].

A recent study [6] of certain 1D models [7] with screening
suppressed by disallowing deposition events which block large
gaps, yielded a surprising conclusion that in these models the
density actually \UN{\it increases} \ away from the substrate
according to the power law

$$ \Delta \rho = \rho(h) - \rho(\infty) \simeq -C h^{-\phi} \;\; ,
\eqno(5)$$

\NI where $C>0$ and $\phi \simeq 0.3$. An interesting question
thus arises: are these models in a universality class different
from KPZ? An alternative is that the KPZ contribution to the
density, (4), is possibly not seen because the added mechanism of
``compactification'' due to suppression of screening, elucidated
in [6], yields the density term (5) with negative exponent $\phi$
\ \UN{\it smaller} \ in absolute value than the KPZ-contribution
exponent.

In this work we report extensive numerical simulations which
clearly place the models under consideration [6-7], to be defined
in detail shortly, within the KPZ universality class. In fact,
the exponent values and scaling form, (1)-(3), are confirmed
quite accurately.

We consider multilayer deposition of $k$-mer ``particles'' on the
linear lattice. The deposition attempts are ``ballistic;'' particles
arrive at a uniform rate per site. The group of those $k$ lattice
sites which are targeted in each deposition attempt is examined to
find the lowest layer $n \geq 1$ such that all the $k$ sites are
empty in that layer (and all layers above it). Note that initially
the substrate is empty, in all the sites and layers $1,2,3,\ldots$.
If the targeted group of sites is in the layer $n=1$, then the
particle is deposited: the $k$ sites become occupied. However, if
the targeted layer is $n>1$, then the deposition attempt is accepted
only provided no gaps of size $k$ or larger are thereby covered in
layer $n-1$. Otherwise the attempt is rejected.

Note that since the layer $n$ is the lowest with all the $k$ sites
empty, then in layer $n-1$, one or two $k$-mer particles partially
or fully cover this group of $k$ sites. Thus within the $k$-span
chosen, the layer $n-1$ can have at most $k-1$ empty sites.
Furthermore, there can be at most 1 continuous sequences of
empty sites which includes one of the ends of the $k$-span in
layer $n-1$. If such a continuous empty-site sequence
actually extends outside the $k$-span (i.e., it includes empty sites
in layer $n-1$ immediately neighboring the chosen
$k$-group of sites), than we disallow deposition if this
``external'' gap is $k$-site or larger. Figures of illustrative
configurations can be found in [6].

Thus we disallow overhangs which would partially or fully block
(screen) those gaps which are large enough to accommodate future
deposition events (in layers $n-1$ or lower). The
final, large-time configuration in each layer contains gaps of at
most $k-1$ consecutive empty sites. However, the gaps can be of
unlimited size in the direction perpendicular to the substrate
(extending from layer to layer). In fact, deposition in lower
layers $1,\ldots,N$ is unaffected by deposition in layers $N+1$ on.
For layer $n=1$, exact solution for the fraction of occupied sites
and for some correlation properties is known [8]; this problem
corresponds to monolayer random sequential adsorption in 1D.

We studied the growth of the interfacial width in this deposition
process. Specifically, we define $L$ as the number of sites in the
lattice (and we use periodic boundary conditions). The Monte Carlo
(MC) time variable $T$ is conveniently defined to have one
deposition attempt per lattice site per unit time. The heights of
the deposit, $h_j$, at sites $j=1,\ldots, L$, were defined as the
number of layers from the substrate to the last occupied layer, at
each lattice site $j$. The r.m.s. width was defined as

$$ W =  \left\langle \ \  \sqrt{{1\over L}\sum_{j=1}^L h^2_j -
\left({1\over L}\sum_{j=1}^Lh_j \right)^2} \ \ \right\rangle
 \;\; , \eqno(6)$$

\NI where the average $\langle \ldots \rangle$ over independent
MC runs was taken after calculating the square root.

Figure~1 presents results for large substrates, $L=2000$.
These data, for $T \leq 200$, were typically averaged over 1000
independent MC runs. There is no visible size effect for $L=2000$.
Thus, relation (1) is replaced by

$$ W \simeq T^{\zeta / z} \;\;\;\;\;\;
\;\;\;\;\;\; (L \to \infty) \;\; , \eqno(7)$$

\NI which corresponds to assuming that the scaling function $F(t)$
behaves according to $\sim t^{\zeta / z}$ for small arguments,

$$ t = TL^{-z} \;\; . \eqno(8)$$

Least-squares fits of the largest-$T$ data indicate that the
exponent in (7) tends to $1/3$. For instance, for data in the range
$150 \leq T \leq 200$, we get $0.383$, $0.345$, $0.316$, $0.304$,
$0.324$, for $k=2,3,5,8,10$, respectively. Based on our analyses, we
propose the estimate

$$ \zeta / z = 0.34 \pm 0.04 \;\; , \eqno(9)$$

\NI clearly excluding values such as $1/2$ or $1/4$ and favoring
the KPZ [1-3] prediction $1/3$. We also checked this estimate for
several data sets taken at $L=1000$ and 1500. The results were
unchanged. For larger $k$, e.g., 10, the onset of the
finite-$L$ saturation can be seen for $L=$O$(1000)$.

Analysis of the finite-$L$ properties was complicated by two
facts. Firstly, to see finite-$L$ saturation, simulations had to
be done for large times. Secondly, we found that the statistical
noise in the data became significant at saturation. Thus averages
over many independent runs were required. We restricted our
extensive MC runs to one $k$ value, $k=3$. This value was favored
because generally, other conditions being equal, the observed
statistical noise became larger as $k$ increased. On the other
hand, the $k=3$ large-$L$ data in Figure~1 yield exponent closer
to $1/3$ than the $k=2$ data, suggesting, possibly, smaller
corrections to the leading scaling behavior.

Figure~2 shows well saturated data for $k=3$ and lattice sizes
$L=20$, 40, 60, 80, 100. These were averaged over typically 10000
MC runs. Shown are also data for $L=300$, averaged over 2000 MC
runs, which have not attained saturation for the largest times
reached in the simulation. The $L=2000$ data from Figure~1 are also
included for comparison.

The spread of the saturation values at
larger $T$ selected to have only the statistical noise, is shown
for $L=20$, 40, 60, 80, 100 in Figure~3. For large times, one
assumes $F(t \to \infty) \simeq {\rm constant}$ \ in (1) so that
the width behaves according to $L^{1/2}$. From least-squares fits
to various data subsets for $L=40$, 60, 80, 100, i.e., excluding
the data for $L=20$ which seem to be too small to reach the true
asymptotic behavior, we propose the estimate

$$ \zeta = 0.49 \pm 0.03   \;\; . \eqno(10)$$

\NI As mentioned earlier, this exponent is the same for various 1D
universality classes and it cannot be used to identify the KPZ
behavior. However, accurate verification of the value $1/2$ suggests
that our data are generally well within the asymptotic regime for
lattice sizes above O$(40)$.

Thus, we also attempted the full scaling data collapse, i.e., we
checked that the quantity

$$ w=WL^{-1/2} \;\; , \eqno(11)$$

\NI when plotted as a function of $t$ defined in (8), is
represented by a unique function $F(t)$; see (1). Of course the
data collapse is exact only in the limit $L \to \infty$ and $T \to
\infty$, with fixed $t$. Figure~4 illustrates the ``collapse'' for
$k=3$, where we used data for $L=80$, 100, 300, 2000, described
earlier. We also included data for $L=1000$ which, together with
the $L=2000$ data, yield the dense portion of the plot for
$w{{}_<\atop{}^\sim}0.3$ (see Figure 4).

All the general expectations on the form of the scaling function
$F(t)$ are qualitatively confirmed by our data. There are several
extensive numerical studies of the KPZ and
other growth-universality classes by scaling data
collapse and tests of universality of quantities derived from
scaling functions similar to $F(t)$; see, for instance, [9-11]. The
quality of our data is comparable to other accurate verifications
of the scaling predictions in 1D, though we found no results in the
literature to allow direct comparison with the scaling-function
data such as Figure~4.

In summary, we found by
extensive MC simulations measuring directly the growing interface
width, that the models with suppressed screening which show unusual
density variation [6] are, in fact, described quite accurately by
the KPZ scaling form [1-3] typical of growing interfaces, with the
appropriate 1D exponent values.

This work was supported in part by the
DFG-Sonder\-forschungs\-bereich 262. One of the authors (P.N.)
wishes to thank the DFG for Heisenberg fellowship.

\NP

\centerline{\bf REFERENCES}{\frenchspacing

\item{[1]} M. Kardar, G. Parisi and Y.C. Zhang, Phys. Rev.
Lett. {\bf 56}, 889 (1986).

\item{[2]} J. Krug and H. Spohn, in {\sl Solids Far from
Equilibrium: Growth, Morphology, Defects}, edited by C.
Godr\`eche (Cambridge University Press, Cambridge, 1991).

\item{[3]} {\sl Dynamics of Fractal
Surfaces}, edited by F. Family and T. Vicsek (World Scientific,
Singapore, 1991).

\item{[4]} J. Krug and P. Meakin, J. Phys. A{\bf 90}, L987 (1990).

\item{[5]} B.D. Lubachevsky, V. Privman and S.C. Roy, Phys. Rev.
E{\bf 47}, 48 (1993).

\item{[6]} P. Nielaba and V. Privman, Phys. Rev. A{\bf 45}, 6099
(1992).

\item{[7]} M.C. Bartelt and V. Privman, J. Chem. Phys. {\bf 93},
6820 (1990).

\item{[8]} J.J. Gonzalez, P.C. Hemmer  and J.S. H{\o}ye, Chem.
Phys. {\bf 3}, 228 (1974).

\item{[9]} J. Krug, P. Meakin and T. Halpin--Healy,
Phys. Rev. A{\bf 45}, 638 (1992).

\item{[10]} M. Schroeder, M. Siegert, D.E. Wolf, J.D. Shore and M.
Plischke, Europhys. Lett. {\bf 24}, 563 (1993).

\item{[11]} M. Siegert and M. Plischke, J. Physique I{\bf 3}, 1371
(1993).

}\NP

\centerline{\bf FIGURE CAPTIONS}

\

\noindent\hang{\bf Fig.~1.}\ \ Data for $k=2,3,5,8,10$ on
substrates of size $L=2000$. For small $T$,
$W_{k=2}<W_{k=3}<\ldots<W_{k=10}$. For large $T$, the relation is
reversed on the average, although the differences,
especially for $k=8$ and 10, are small and fluctuate in sign due
to statistical noise. Solid line illustrates slope $1/3$.

\noindent\hang{\bf Fig.~2.}\ \ Data
for $k=3$ on substrates of sizes $L=20$, 40, 60, 80, 100, 300,
2000. For fixed $T$, $W(L)$-values monotonically increase with $L$.
Solid line illustrates slope $1/3$.

\noindent\hang{\bf Fig.~3.}\ \ Large-$T$ data
for $k=3$ on substrates of sizes $L=20$, 40, 60, 80, 100,
illustrating the $L$-dependence of the saturation values (with the
statistical noise). Solid line illustrates slope $1/2$.

\noindent\hang{\bf Fig.~4.}\ \ Scaling data collapse according to
equation (1), for $k=3$. The scaled width $w$ is plotted as a
function of the scaled time $t$, for substrate sizes $L=80$, 100,
300, 1000, 2000.

\bye